\documentclass[sigconf]{acmart}

\renewcommand\footnotetextcopyrightpermission[1]{} 
\usepackage[T1]{fontenc}
\usepackage{times}%
\usepackage{amsmath}
\usepackage{amsthm}
\newtheorem{mydef}{Definition}
\newtheorem{mythm}{Theorem}
\usepackage[utf8]{inputenc}
\usepackage{amsmath}
\usepackage{amsfonts}
\usepackage{amssymb}
\usepackage{dblfloatfix}
\usepackage{colortbl}
\usepackage{multirow}
\usepackage{algorithm}
\usepackage{scalefnt}
\usepackage{algorithmic}
\usepackage{mathtools}
\usepackage[font=small,labelfont=bf]{caption}
\usepackage[normalem]{ulem}
\usepackage{pgfplots}
\usetikzlibrary{matrix}
\usepgfplotslibrary{groupplots}
\pgfplotsset{compat=newest}
\pgfplotsset{compat=newest,
legend style={font=\footnotesize},
label style={font=\footnotesize},
tick label style={font=\footnotesize},
title style={font=\footnotesize}}
\usepackage{tikz}
\DeclareMathOperator*{\minimize}{minimize}
\DeclareMathOperator*{\argmax}{argmax}
\usepackage{fancyhdr}
\usepackage{natbib}
\usepackage{amsmath}
\usepackage{dcolumn}
\usepackage{graphics}

\begin{document}

\title{Fault Tolerance of Neural Networks in Adversarial Settings}

\author{Vasisht Duddu$^1$, N. Rajesh Pillai$^2$, D. Vijay Rao$^3$, Valentina E. Balas$^4$}
\affiliation{\institution{$^1$ Indraprastha Institute of Information Technology, Delhi, India\\$^2$  \large Scientific Analysis Group, Delhi, India\\ $^3$  \large Institute for Systems Studies and Analyses, Delhi, India\\ $^4$ \large Aurel Vlaicu University of Arad, Arad, Romania}}
\email{vduddu@tutamail.com, rpillai@sag.drdo.in, vijayrao@issa.drdo.in, valentina.balas@uav.ro}

\begin{abstract}
Applications using Artificial Intelligence techniques demand a thorough assessment of different aspects of trust, namely, data and model privacy, reliability, robustness against adversarial attacks, fairness, and interpretability.
While each of these aspects has been extensively studied in isolation, an understanding of the trade-offs between different aspects of trust is lacking.
In this work, the trade-off between fault tolerance, privacy, and adversarial robustness is evaluated for Deep Neural Networks, by considering two adversarial settings under security and a privacy threat model.
Specifically, this work studies the impact of training the model with input noise (Adversarial Robustness) and gradient noise (Differential Privacy) on Neural Network's fault tolerance.
While adding noise to inputs, gradients or weights enhances fault tolerance, it is observed that adversarial robustness lowers fault tolerance due to increased overfitting.
On the other hand, ($\epsilon_{dp},\delta_{dp}$)-Differentially Private models enhance the fault tolerance, measured using generalisation error, which theoretically has an upper bound of $e^{\epsilon_{dp}} - 1 + \delta_{dp}$.
This novel study of the trade-offs between different aspects of trust is pivotal for training trustworthy Machine Learning models.
\end{abstract}

\maketitle
\thispagestyle{fancy}

\section{Introduction}

\noindent There is a growing reliance on Artificial Intelligence (AI) techniques in safety-critical real-time applications with high-stake decision-making such as autonomous vehicles, criminal justice, and healthcare.
These applications demand satisfying different aspects of trust: fairness among disparate groups, the privacy of individuals in the training data, robustness against adversarially perturbed inputs, fault tolerance for safety and model interpretability.
Training the models to incorporate and optimize for all the aspects of trust is difficult and hence, it is crucial to understand the trade-offs between different aspects of trust for designing efficient Pareto-optimal solutions for trustworthy AI systems.
Prior research has indicated that privacy and explainability \cite{shokri2019privacy}, adversarial robustness and membership privacy \cite{song2019privacy} are at odds.
On the other hand, other aspects of trust go hand in hand: fairness and differential privacy \cite{pmlr-v97-jagielski19a}, adversarial robustness and explainability \cite{pmlr-v97-etmann19a}.
However, the impact on fault tolerance due to robust and private models has not been explored yet.
To address this requirement, this research analyses the impact of training Machine Learning models, specifically Deep Neural Networks, for adversarial robustness (security) and differential privacy on the model's fault tolerance (reliability).

Fault Tolerance is a crucial property for Neural Networks to ensure reliable computation for a long duration with graceful degradation over time.
Typically, well generalized models have the parameters with low variance ensuring equal computational weight to all nodes in the network \cite{duddu2019adversarial}.
Hence, the loss of some of the nodes can be compensated by other nodes without a significant loss in performance \cite{7862272}\cite{712162}\cite{105414}.
Practically, this is achieved by adding noise during training to the inputs, gradients or the weights.
The noise added to the inputs can be modeled as Tikhonov regularization which enhances the generalization \cite{Bishop:1995:TNE:211171.211185}.

However, in an adversarial setting under a security threat model, carefully crafted \textit{imperceptible} noise can be added to input images by an adversary to force the model to misclassify the image, violating the integrity of the model prediction.
The design of Neural Networks within such an adversarial setting requires training on inputs with adversarial noise to ensure robustness against adversary's worst-case perturbation.
The goal of this work, in this setting, is to address the following research question,
\begin{quote}
\textit{What is the impact of adversarially robust (input noise) training on fault tolerance?}
\end{quote}

Alternatively, in an adversarial setting under a privacy threat model, an adversary performs inference attacks to identify training data attributes or membership details for sampled data point \cite{Ganju:2018:PIA:3243734.3243834}\cite{7958568}\cite{ndss19salem}.
This poses a serious privacy risk for sensitive training data such as financial and medical records, personal photos, location history, and user preferences.
Differential Privacy provides a provable guarantee on the maximum privacy leakage by making the data points indistinguishable using gradient noise during training \cite{noauthororeditor}\cite{Abadi:2016:DLD:2976749.2978318}.
In such a setting, this work addresses the following research question,
\begin{quote}
\textit{What is the impact of training models with Differential Privacy (gradient noise) on fault tolerance?}
\end{quote}

\textbf{Main Contributions.} This work makes the following novel contributions:
\begin{itemize}
\item Evaluate the fault tolerance of provably robust Neural Networks and compare them with the theoretical equivalent Tikhonov regularized model.
\item Evaluate the fault tolerance of Differentially Private models under a privacy threat model and compare it with regularised and naturally (without noise) trained models.
\item Prove theoretically that generalization error, used to measure fault tolerance, has an upper bound of $e^{\epsilon_{dp}}-1+\delta$ on training the model using ($\epsilon_{dp},\delta_{dp}$)-Differential Privacy, thus giving provable guarantees on fault tolerance metric.
\end{itemize}

The performance of the training algorithms is evaluated using two common benchmarking datasets: CIFAR10 and FashionMNIST, with different Neural Network architectures.
It is observed that adversarial robustness and fault tolerance are at odds with each other, i.e, training a model with adversarial input noise results in overfitting which lowers fault tolerance.
On the other hand, noise added for ($\epsilon_{dp}, \delta_{dp}$)-Differential Privacy is an alternate approach for enhancing fault tolerance, while guaranteeing privacy, with a theoretical bound on the generalization error in terms of the privacy parameters: $\epsilon_{dp}, \delta_{dp}$.
To the best of our knowledge, this is the first work that evaluates the fault tolerance of adversarially robust and differentially private Deep Neural Networks.
Such an analysis is crucial for a unified framework for trustworthy Machine Learning combining security, privacy, and reliability for real-world deployment.

\begin{table}[!htb]
\caption{Variable notations for Reliability, Adversarial Robustness and Differentially Privacy.}
\centering
\renewcommand\arraystretch{1.5}
\fontsize{6.7pt}{6.7pt}\selectfont
\resizebox{\columnwidth}{!}{
\begin{tabular}{|c|c|}
\hline
\textbf{Symbol} & \textbf{Description}\\
\hline
\multirow{1}{*}{$\mathcal{F}$} & \multirow{1}{*}{Target Machine Learning Model} \\
\multirow{1}{*}{$\mathcal{D}_{train}$} & \multirow{1}{*}{Training Dataset} \\
\multirow{1}{*}{$\mathcal{D}_{test}$} & \multirow{1}{*}{Testing Dataset} \\
\multirow{1}{*}{$\mathcal{X}$} & \multirow{1}{*}{Space of Input Data Points} \\
\multirow{1}{*}{$\mathcal{Y}$} & \multirow{1}{*}{Space of Output Labels} \\
\multirow{1}{*}{$\textbf{(x,y)}$} & \multirow{1}{*}{Data Sample with input and output label} \\
\multirow{1}{*}{$\textbf{P(x,y)}$} & \multirow{1}{*}{Data Distribution over all samples} \\
\multirow{1}{*}{$\mathcal{L}$} & \multirow{1}{*}{Loss Function for Training} \\
\hline
\multirow{1}{*}{\textbf{Fault Tolerance}} & \multirow{1}{*}{} \\
\multirow{1}{*}{$\epsilon_{ft}$} & \multirow{1}{*}{Fault Tolerance Bound} \\
\multirow{1}{*}{$\mathcal{N}$} & \multirow{1}{*}{Ideal Neural Network without Faults} \\
\multirow{1}{*}{$\mathcal{N}_{fault}$} & \multirow{1}{*}{Neural Network model with Faults} \\
\multirow{1}{*}{$\mathcal{H}_{N}$} & \multirow{1}{*}{Trained Ideal (faultless) Neural Network} \\
\multirow{1}{*}{$\mathcal{H}_{N_{fault}}$} & \multirow{1}{*}{Trained Faulty Neural Network} \\
\multirow{1}{*}{$\mathbb{G}_{error}$} & \multirow{1}{*}{Generalization Error; Metric for measuring fault tolerance} \\
\hline
\multirow{1}{*}{\textbf{Adversarial Robustness}} & \multirow{1}{*}{} \\
\multirow{1}{*}{$\epsilon_{adv}$} & \multirow{1}{*}{Adversarial Noise Bounds} \\
\multirow{1}{*}{$\Delta$} & \multirow{1}{*}{Set of Adversarial Noise Values bounded by $\epsilon_{adv}$} \\
\multirow{1}{*}{$\delta_{adv}$} & \multirow{1}{*}{Value of Adversarial Noise in $\Delta$} \\
\hline
\multirow{1}{*}{\textbf{Differential Privacy}} & \multirow{1}{*}{} \\
\multirow{1}{*}{$\epsilon_{dp}$} & \multirow{1}{*}{Differential Privacy Leakage Bound} \\
\multirow{1}{*}{$\delta_{dp}$} & \multirow{1}{*}{Differential Privacy failure probability} \\
\multirow{1}{*}{$\mathbb{E}_{train}$} & \multirow{1}{*}{Error on training set} \\
\multirow{1}{*}{$\mathbb{E}_{test}$} & \multirow{1}{*}{Error on test set} \\
\multirow{1}{*}{$\mathbb{D}'$} & \multirow{1}{*}{Dataset after adding/removing single data point} \\
\hline
\end{tabular}
}
\label{tab:notations}
\end{table}

\section{Background}\label{background}

\subsection{Fault Tolerance in Neural Networks}

\begin{mydef}
A Neural Network $\mathcal{N}$ performing computations $\mathcal{H}_{\mathcal{N}}$ is said to be fault tolerant if the computation performed by a faulty network $\mathcal{H}_{\mathcal{N}_{fault}}$ is close to $\mathcal{H}_{\mathcal{N}}$.
Formally, a Neural Network is $\epsilon$ fault tolerant if,
\begin{equation}\label{eft}
\left\lVert \mathcal{H}_{\mathcal{N}} (\mathcal{X}) - \mathcal{H}_{\mathcal{N}_{fault}} (\mathcal{X}) \right\rVert \leq \epsilon_{ft}
\end{equation}
for $\epsilon_{ft} > 0$ and $\mathcal{X} \in \mathcal{D}$.
\end{mydef}

\textbf{Fault Tolerance Metric.} Improving the generalization results in enhancing the fault tolerance and vice versa \cite{article}.
Hence, capturing the overfitting of the model provides a way to compare the relative fault tolerance of multiple models.
This has been extensively used in literature to measure fault tolerance of Neural Networks \cite{8561200} \cite{duddu2019adversarial}.
Formally, given a dataset $\mathcal{D}_{test}$ such that $\mathcal{D}_{train}  \cap  \mathcal{D}_{test} = \phi$, the accuracy of the model is estimated on the training set ($R_{train}$) and on the testing set ($R_{test}$).
The generalization error is given by the difference between training accuracy and the testing accuracy,
\begin{equation}
\mathcal{G}_{err} = \mathcal{R}_{train} - \mathcal{R}_{test}
\end{equation}
This gives the estimate of overfitting in the Neural Network ($R_{train} > R_{test}$), i.e, higher the generalization error more the overfitting.
This estimate of fault tolerance is used for the evaluation of different Neural Networks throughout the paper.

\textbf{Fault Model.} In this work, the faults occurring in the hardware are simulated in the form of stuck at ``0" errors during the Neural Network computation.
Further, multiple faults can occur simultaneously for which the performance degradation is measured using test accuracy.
The faults are simulated in two ways: firstly, these faults can manifest in the form of random node crashes in the Neural Network due to which the output of the node is forced to zero.
Secondly, the parameters of the Neural Networks can be stuck at ``0" which includes weights in the case of Multilayer Perceptron, and filter and kernel values in case of Convolutional Neural Networks.
This is a common fault model frequently used in evaluating the reliability of systems \cite{duddu2019adversarial}.

\textbf{Related Work.} A widely used approach for enhancing fault tolerance is to penalize large values of the parameters using a regularization function \cite{7862272}\cite{8561200}.
Alternatively, constraint optimisation approaches using minimax constraint \cite{105414} as well as quadratic programming \cite{712162} can be used for small networks.
Unlike simple regularization functions, unsupervised pre-training of the initial network layers followed by supervised fine-tuning can significantly enhance the fault tolerance \cite{duddu2019adversarial}.
Traditional techniques to enhance reliability such as additional redundancy by adding nodes and synapses provides partial fault tolerance \cite{363479}.
Further, reliability of axonal transport has been explored using Hammock Networks \cite{7980404}\cite{8340051}.
Detection of faults and enhancing tolerance in software has been explored for fuzzy control systems \cite{Jin:2014:AFI:2596370.2596386}.

\subsection{Adversarial Robustness}

Within the adversarial setting with a security adversary, the problem of adversarial robustness is modeled as a game between the attacker and defender with conflicting interests.
Here, the adversary wants to force the target model to misclassify by adding carefully calculated noise to input, while the defender wants to train the model to defend against such inputs with adversarial noise \cite{Duddu}.

\textbf{Attacker Knowledge.} In this work, the adversary has no knowledge about the target model.
In other words, the adversary has remote access to the target black-box model and can query the model and receive corresponding predictions through an API.
This is typically the black box setting seen in Machine Learning as a Service.

\textbf{Attacker Goal.} The goal of the adversary is to find the noise to maximize the loss of the target model ($F_{\theta}()$) and force the model to misclassify the perturbed input.
Formally, given an input sampled from the underlying data distribution $x \sim P(X, Y)$, the adversary computes the worst-case adversarial noise $\delta_{adv}$ to maximize the loss ($l$) between predicted output and true output $y$,
\begin{equation}\label{untarg}
\delta^\star(x)_{adv} = \argmax_{\|\delta_{adv}\| \leq \epsilon_{adv}} \ell(F_\theta(x + \delta_{adv}), y)
\end{equation}
subject to a bound on the perturbation computed using parameter $\epsilon_{adv}$,
\begin{equation}
\Delta = \{\delta_{adv}: ||\delta_{adv}||_p \leq \epsilon_{adv}\}
\end{equation}
The optimization is subjected to the condition that the noise is imperceptible by restricting $\delta$ within a perturbation region $\Delta$ defined by a $l_p$ norm, more commonly $l_{\infty}$ \cite{madry2018towards}.

\textbf{Defender Strategy.} To defend against the worst attack possible, the model is trained using adversarial inputs as part of the training data.
Formally, this empirical defense can be modeled as a minimax optimization problem given below,
\begin{equation}\label{minmax}
\minimize_\theta \left(\frac{1}{|D|} \sum_{x,y \in D} \max_{\|\delta\| \leq \epsilon} \ell(F_\theta(x + \delta), y)\right)
\end{equation}
Here, instead of minimising the expected loss over the data points sampled from the distribution, the optimisation minimises the worst case loss over the data with adversarial noise.
In other words, the defender minimises the loss corresponding to the worst case adversarial attack.

In this work, TRADES algorithm is considered as a defense since it provides provable bounds against adversarial examples \cite{pmlr-v97-zhang19p}.
TRADES algorithm decomposes the prediction error for adversarial example as the sum of natural classification error and the boundary error to provide a tight differentiable upper bound.
This defense minimizes the maximum Kullback-Leibler (KL) Divergence between the output prediction corresponding to a benign sample ($X$) and adversarial sample ($X_{adv} \leftarrow X + \delta_{adv}$).
This is used to generate adversarial examples within the inner maximization.
\begin{multline}
\min_{\theta} ( \mathbb{E} [l(\theta, F(X), Y)] + \max_{X_{adv} \in \mathcal{\Delta}(X,\epsilon)} d_{kl}(F(X),F(X_{adv})) )
\end{multline}

\begin{algorithm}[!htb]
\caption{Adversarial Training by adding noise to inputs}\label{alg:advtrain}
\textbf{Input:} $D_{train}$ = \{$ (x_i,y_i), \cdots, (x_N,y_N) $\} and Loss: $L(\theta_t,x_i)$\\
\textbf{Input:} $\mathbb{A}$: Algorithm to generate adversarial noise
\begin{algorithmic}[1]
\FOR {each epoch}
\STATE {Sample a random batch $B \in D_{train}$}
\FOR {each $(x_i,y_i)$ $\in B$}
\STATE {\textbf{Compute Adversarial Noise:}\\ $x_{adv}^* \leftarrow A(x_i,y_i)$}\\
\STATE {\textbf{Compute gradient on adversarial inputs:}\\ $\frac{\partial L_i}{\partial \theta_t} \leftarrow \nabla_{\theta_t} L(\theta_t, x_{adv}^*)$}\\
\STATE {\textbf{Update $\theta$ using Gradient Descent:}\\ $\theta_{t+1} \leftarrow \theta_{t} - \alpha \frac{\partial L_i}{\partial \theta_t}$}
\ENDFOR
\STATE{\textbf{Output:} Parameters $\theta$ of trained model with adversarial robustness}
\ENDFOR
\end{algorithmic}
\end{algorithm}

\textbf{Related Work.} While only the defense algorithm with tight upper bound and provable robustness guarantees is considered in this work (TRADES), other empirical approaches use Projected Gradient Descent \cite{madry2018towards} and Wasserstein norm \cite{sinha2018certifiable} to ensure robustness.
Alternatively, verification based defenses use function transformations to compute the worst-case loss to express the adversarial perturbations \cite{pmlr-v80-mirman18b}.

\subsection{Differential Privacy}

Differential Privacy is the de facto privacy standard that provides a strong privacy definition with provable bounds on information leaked by a mechanism in terms of the privacy budget $\epsilon_{dp}$ \cite{noauthororeditor}.
The output of a randomized mechanism should not allow the adversary to learn any more about an individual in the dataset than that could be learned via the same analysis without the individual in the dataset.
In this sense, this definition of privacy captures the individual's membership in the dataset.

\begin{mydef}
For a randomized mechanism $\mathcal{M} : \mathcal{X} \rightarrow \mathcal{Y}$ is $(\epsilon_{dp},\delta_{dp})$ differentially private on two neighbouring datasets $\mathcal{D}$ and $\mathcal{D'}$ differing by an individual data point, then for all outputs $\mathcal{O}$ $\subseteq$ $\mathcal{Y}$,
\begin{equation}
\mathcal{P}[\mathcal{M(D)} \in \mathcal{O}] \leq e^{\epsilon_{dp}} \mathcal{P}[\mathcal{M(D')} \in \mathcal{O}] + \delta_{dp}
\end{equation}
\end{mydef}
Here, the parameters $\epsilon_{dp}$ is considered as the privacy budget and $\delta_{dp}$ is considered as the failure probability \cite{10.1007/978-3-540-79228-4}.

A tighter and accurate estimation of the privacy loss can be computed using the Renyi Differential Privacy \cite{8049725} which uses the Renyi divergence metric $\mathcal{D}_{\alpha}$ which applies to any moment of the privacy loss random variable.
\begin{mydef}
For a randomized mechanism $\mathcal{M} : \mathcal{X} \rightarrow \mathcal{Y}$ is $\epsilon_{dp}$-Renyi differentially private of the order $\alpha$, on two neighbouring datasets $\mathcal{D}$ and $\mathcal{D'}$ differing by a individual data point, then for all outputs $\mathcal{O}$ $\subseteq$ $\mathcal{Y}$,
\begin{equation}
\mathcal{D}_{\alpha}(\mathcal{M(D)}  \mid  \mid \mathcal{M(D')}) \leq \epsilon
\end{equation}
\end{mydef}

\textbf{Attacker Strategy.} The goal of the attacker within the privacy setting is to use inference attacks to leak training data details resulting in privacy violations where the training data is sensitive.
Membership inference attacks infer whether a given data point was used in the training data or not based on the difference in model performance on training data and testing data \cite{7958568}.
On the other hand, attribute inference attacks extract particular features of the training data \cite{Fredrikson:2015:MIA:2810103.2813677} or reconstruct the entire training data \cite{DBLP:journals/corr/abs-1904-01067}.
Another class of attacks exploits the memorization capacity of the model to infer the sensitive attributes in the data by querying the model \cite{236216}.

\textbf{Defender Strategy.} The defender, in order, to reduce the success of the inference attacks utilizes the notion of differential privacy to train the model with provable privacy leakage guarantees.
To this extent, the defender samples a noise from either a Laplace or Gaussian distribution proportional to the sensitivity of the model ($S$).

\begin{equation}
\mathcal{M}(D) \simeq \mathcal{H}_{\mathcal{N}} (\mathcal{X}) + \mathcal{N}(0,S^2.\sigma^2)
\end{equation}

In case of Deep Neural Networks, this noise is added to the gradients of the model and mitigates several of the attacker's inference strategies, however, at the cost of utility (performance) \cite{236254}.

\textbf{Related Work.} To preserve privacy against membership inference attacks, several empirical defenses exist.
For instance, the adversary's inference attack can be modeled as a minimax optimization problem, where the target model is trained to minimize the adversary's best attack \cite{Nasr:2018:MLM:3243734.3243855}.
Another line of defense is to add noise to the output of the model, force the inference attack machine learning model to misclassify while ensuring the utility does not degrade \cite{jia2019memguard}.
While these approaches are empirical, all of them face utility privacy trade-off and do not provide a theoretical guarantee on the maximum leakage of the model.
Hence, in this work, Differential Privacy based private training is used since it provides provable guarantees on the leakage of the model about the training data \cite{Abadi:2016:DLD:2976749.2978318}.
An alternative Differential Privacy based framework uses a teacher-student ensemble approach \cite{papernot2018scalable}.

\begin{algorithm}[!htb]
\caption{Differentially Private Training by adding noise to gradients during Backpropagation}\label{alg:privsgd}
\textbf{Input:} $D_{train}$ = \{$ (x_i,y_i), \cdots, (x_N,y_N) $\} and Loss function: $L(\theta_t,x_i)$
\begin{algorithmic}[1]
\FOR {each epoch}
\STATE {Sample a random batch $B \in D_{train}$}
\FOR {each $(x_i,y_i)$ $\in B$}
\STATE {\textbf{Compute gradient:}\\ $\frac{\partial L_i}{\partial \theta_t} \leftarrow \nabla_{\theta_t} L(\theta_t, x_i)$}\\
\STATE {\textbf{Gradient Clipping:}\\ $g_t(x_i) \leftarrow \frac{\partial L_i}{\partial \theta_t} / \max\big(1, \frac{\|\frac{\partial L_i}{\partial \theta_t} \|_2}{C}\big)$}\\
\STATE {\textbf{Add Noise:}\\ $g'_t \leftarrow \frac{1}{L}\left( \sum_i g_t(x_i) + \mathcal{N}(0, \sigma^2 C^2 I)\right)$}
\STATE {\textbf{Update $\theta$ using Gradient Descent:}\\ $\theta_{t+1} \leftarrow \theta_{t} - \alpha g_t'$}
\ENDFOR
\STATE{\textbf{Output:} Parameters $\theta$ of model trained using ($\epsilon,\delta$)-Differential Privacy.}
\ENDFOR
\end{algorithmic}
\end{algorithm}

\section{Experiment Setup}

The code for training using TRADES adversarial training algorithm is based on the official source code from the authors\footnote{https://github.com/yaodongyu/TRADES}.
The official code from Tensorflow Privacy Library\footnote{https://github.com/tensorflow/privacy} for Differentially Private training is adapted to different architectures and datasets.

\subsection{Datasets}

The evaluation and training of adversarially robust and differentially private models are done on two major benchmarking datasets, namely, FashionMNIST and CIFAR10.

\textbf{FashionMNIST.} The dataset is similar to the MNIST dataset and consists of a training set of 60,000 examples and a test set of 10,000 examples.
Each data sample is a 28$\times$28 grayscale image associated with a label from 10 classes such as boots, shirt, bag and so on.

\textbf{CIFAR10.} The CIFAR-10 dataset consists of 60000 32x32 colour images in 10 classes, with 6000 images per class.
There are 50000 training images and 10000 test images.

\subsection{Architectures}

\begin{table}[!htb]
\caption{Architectures for CIFAR10 and FashionMNIST datasets.}
\centering
\renewcommand\arraystretch{1.5}
\fontsize{6.7pt}{6.7pt}\selectfont
\resizebox{\columnwidth}{!}{
\begin{tabular}{|c|c|c|}
\hline
\textbf{CIFAR10 Architecture} & \textbf{FMNIST CNN 1} & \textbf{FMNIST CNN 2}\\
\hline
Convolution  32 (3x3)(1) & Convolution  20 (5x5)(1) & Convolution  16 (8)(1)\\
Convolution  32 (4x4)(2) & MaxPool (2) (2) & MaxPool (2) (1)\\
Convolution  32 (3x3)(1) & Convolution  50 (5x5)(1) & Convolution  32 (4)(2)\\
Convolution  64 (3x3)(2) & MaxPool (2) (2) & MaxPool (2) (1)\\
Convolution  64 (3x3)(1) & Dense 500 & Dense 32\\
Convolution 128 (3x3)(2) & Dense 10 & Dense 10\\
Convolution 128 (3x3)(1) & &\\
Convolution 256 (3x3)(2) & & \textbf{FMNIST DNN}\\
Convolution 256 (3x3)(1) & & Dense 512\\
Dense 512 & & Dense 512\\
Dense 512 & & Dense 10\\
Dense 10 & &\\
\hline
\end{tabular}
}
\label{tab:architectures}
\end{table}

For the TRADES robust training algorithm on CIFAR10 dataset, an Neural Network architecture with nine convolutional layers followed by fully connected layers is used.
In the case of FashionMNIST dataset with adversarial robust training, the FMNIST CNN 2 architecture based on LeNet architecture is used.
For Differential Private training, FMNIST CNN 1 architecture is used with minor differences in hyperparameters to the FMNIST CNN 2 architecture.
Further, for evaluating on a simple Multilayer Perceptron Network, a Neural Network with two hidden layers of sizes 512 nodes each is used.
The details of the exact architectures used in the experiments are given in Table 2.

\section{Fault Tolerance and Adversarial Robustness}\label{ft_advtrain}

\subsection{Input Noise as a Regularizer}

Adding noise to inputs has been shown to provide a regularization effect that theoretically is equivalent to Tikhonov regularization \cite{Bishop:1995:TNE:211171.211185}.
To understand the effect on input noise to generalization, a simple architecture of 500 hidden layers is considered for a binary classification problem and differentiating two types of circles\footnote{sklearn.datasets.make\_circles}.
Here, two types of noise are considered, namely, additive Gaussian noise and the multiplicative Gaussian noise which are commonly used for enhancing the fault tolerance of Neural Networks by improving the generalization \cite{5446319} \cite{105415} \cite{155944}.
This indicates that adding noise enhances for fault tolerance for small noise values beyond which the model overfits and the performance degrades.

\begin{table}[!htb]
\begin{center}
\caption{Generalization Error for additive gaussian noise and multiplicative gaussian noise added to inputs.}
\renewcommand\arraystretch{1.5}
\fontsize{6.7pt}{6.7pt}\selectfont
\resizebox{\columnwidth}{!}{
\begin{tabular}{ |c|c|c|c| }
\hline
 \textbf{$\sigma$} & \textbf{Training Accuracy} & \textbf{Testing Accuracy} & \textbf{Generalization Error}\\
 \hline
 \hline
\multicolumn{4}{|c|}{\underline{\textbf{No Noise}}} \\
\textbf{0} & \cellcolor{gray!20}100.00\% & \cellcolor{gray!20}75.7\% & \cellcolor{gray!20}24.3\%\\
\hline
\hline
  \multicolumn{4}{|c|}{\underline{\textbf{Additive Gaussian Noise}}} \\
\textbf{0.01}  & \cellcolor{green!20}100.00\% & \cellcolor{green!20}77.1\% & \cellcolor{green!20}22.9\% \\
\textbf{0.1}  & \cellcolor{red!20}93.3\% & \cellcolor{red!20}67.1\% &  \cellcolor{red!20}26.2\% \\
\textbf{0.5} & \cellcolor{red!20}73.3 & \cellcolor{red!20}44.3 & \cellcolor{red!20}29.0\%\\
\hline
\hline
\multicolumn{4}{|c|}{\underline{\textbf{Multiplicative Gaussian Noise}}} \\
\textbf{0.03}  & \cellcolor{green!20}100.00\% & \cellcolor{green!20}77.1\% & \cellcolor{green!20}22.9\% \\
\textbf{0.1}  & \cellcolor{red!20}96.7\% & \cellcolor{red!20}71.4\% &  \cellcolor{red!20}25.3\% \\
\textbf{0.5} & \cellcolor{red!20}80.0 & \cellcolor{red!20}48.6 & \cellcolor{red!20}31.4\%\\
 \hline
\end{tabular}
}
\end{center}
\end{table}

In the case of additive noise, on adding a small noise of a standard deviation of 0.01, the generalization performance on the binary classification problem improves (Table 3).
However, on increasing the values of the standard deviation, the model starts to overfit the noisy training data and the performance starts to decline.
A similar phenomenon is observed on training models in the presence of worst-case adversarial noise as shown in the subsequent sections.

\subsection{Adversarial Noise}

\begin{figure}[!htb]
\begin{center}
\resizebox{0.8\columnwidth}{!}{%
\begin{tikzpicture}[font=\tiny]
\begin{axis}[
legend style={font=\scriptsize},
legend pos =  north west,
legend entries={Natural, Trades, DistAdv},
ylabel={Generalisation Error},
xlabel={Epochs},
xmin=0, ymin=0,
xmax=150,
every axis plot/.append style={ultra thick},
grid=major
]
\addplot[
    color=black,
    solid,
    smooth
    ]
    coordinates {
    (0,-0.69)
  (1,-0.44)
  (2,-1.41)
  (3,0.11)
  (4,-0.40)
  (5,0.40)
  (6,-0.17)
  (7,0.39)
  (8,0.82)
  (9,0.93)
  (10,1.27)
  (11,1.36)
  (12,1.50)
  (13,1.67)
  (14,2.14)
  (15,2.45)
  (16,2.48)
  (17,2.62)
  (18,2.76)
  (19,3.27)
  (20,3.57)
  (21,3.44)
  (22,2.96)
  (23,3.58)
  (24,3.54)
  (25,3.97)
  (26,3.55)
  (27,4.29)
  (28,4.59)
  (29,4.79)
  (30,4.67)
  (31,4.41)
  (32,4.55)
  (33,4.73)
  (34,4.17)
  (35,5.05)
  (36,5.47)
  (37,4.92)
  (38,5.31)
  (39,5.17)
  (40,4.99)
  (41,5.07)
  (42,5.72)
  (43,6.01)
  (44,6.46)
  (45,5.85)
  (46,6.44)
  (47,4.85)
  (48,6.54)
  (49,5.81)
  (50,6.17)
  (51,6.48)
  (52,6.01)
  (53,6.80)
  (54,6.42)
  (55,5.95)
  (56,6.43)
  (57,6.27)
  (58,6.80)
  (59,6.30)
  (60,6.53)
  (61,7.10)
  (62,6.70)
  (63,7.12)
  (64,7.42)
  (65,6.41)
  (66,7.27)
  (67,6.78)
  (68,7.18)
  (69,7.16)
  (70,7.01)
  (71,7.25)
  (72,8.21)
  (73,7.15)
  (74,7.51)
  (75,8.14)
  (76,7.19)
  (77,7.48)
  (78,7.87)
  (79,7.67)
  (80,8.19)
  (81,7.93)
  (82,7.90)
  (83,7.80)
  (84,8.05)
  (85,8.50)
  (86,7.94)
  (87,7.88)
  (88,7.21)
  (89,8.13)
  (90,7.86)
  (91,7.82)
  (92,7.99)
  (93,8.16)
  (94,8.29)
  (95,7.82)
  (96,7.99)
  (97,8.43)
  (98,8.48)
  (99,8.54)
  (100,8.40)
  (101,8.74)
  (102,8.18)
  (103,8.74)
  (104,8.77)
  (105,8.64)
  (106,8.72)
  (107,8.67)
  (108,8.48)
  (109,8.86)
  (110,8.86)
  (111,8.52)
  (112,9.11)
  (113,8.24)
  (114,8.27)
  (115,9.03)
  (116,8.46)
  (117,8.93)
  (118,8.93)
  (119,8.60)
  (120,8.70)
  (121,8.95)
  (122,9.15)
  (123,9.20)
  (124,9.17)
  (125,8.70)
  (126,8.59)
  (127,9.05)
  (128,8.79)
  (129,9.44)
  (130,9.27)
  (131,9.03)
  (132,9.29)
  (133,9.63)
  (134,8.94)
  (135,9.56)
  (136,9.13)
  (137,9.23)
  (138,9.87)
  (139,9.67)
  (140,9.53)
  (141,8.71)
  (142,9.53)
  (143,9.54)
  (144,9.13)
  (145,9.37)
  (146,9.31)
  (147,9.30)
  (148,9.26)
    };
\addplot[
      color=red,
      solid,
      smooth
    ]
    coordinates {
    (0,-1.15)
  (1,-2.15)
  (2,-2.98)
  (3,-1.91)
  (4,-0.06)
  (5,-1.25)
  (6,-1.14)
  (7,-0.05)
  (8,-0.13)
  (9,1.16)
  (10,1.19)
  (11,0.90)
  (12,1.34)
  (13,0.49)
  (14,1.24)
  (15,1.39)
  (16,1.79)
  (17,2.15)
  (18,2.19)
  (19,2.56)
  (20,3.46)
  (21,2.34)
  (22,3.75)
  (23,3.56)
  (24,3.72)
  (25,3.96)
  (26,3.96)
  (27,3.23)
  (28,4.73)
  (29,4.23)
  (30,4.36)
  (31,4.00)
  (32,4.23)
  (33,4.47)
  (34,3.73)
  (35,5.28)
  (36,4.39)
  (37,5.01)
  (38,5.61)
  (39,5.74)
  (40,5.35)
  (41,5.88)
  (42,6.11)
  (43,6.55)
  (44,6.38)
  (45,5.76)
  (46,6.10)
  (47,6.65)
  (48,6.87)
  (49,7.43)
  (50,7.19)
  (51,7.07)
  (52,7.50)
  (53,7.49)
  (54,7.60)
  (55,7.69)
  (56,8.01)
  (57,7.65)
  (58,8.37)
  (59,8.11)
  (60,7.96)
  (61,7.30)
  (62,8.65)
  (63,8.54)
  (64,8.43)
  (65,9.24)
  (66,9.57)
  (67,9.79)
  (68,10.36)
  (69,9.63)
  (70,10.07)
  (71,9.90)
  (72,9.82)
  (73,10.00)
  (74,10.03)
  (75,10.13)
  (76,10.18)
  (77,10.58)
  (78,11.40)
  (79,10.87)
  (80,10.37)
  (81,10.66)
  (82,11.03)
  (83,11.21)
  (84,11.67)
  (85,11.44)
  (86,11.37)
  (87,10.75)
  (88,10.87)
  (89,11.34)
  (90,10.84)
  (91,11.12)
  (92,11.72)
  (93,11.71)
  (94,11.92)
  (95,12.54)
  (96,11.83)
  (97,12.77)
  (98,11.81)
  (99,11.80)
  (100,11.82)
  (101,12.57)
  (102,13.48)
  (103,12.80)
  (104,12.55)
  (105,12.80)
  (106,12.70)
  (107,13.24)
  (108,12.93)
  (109,13.40)
  (110,13.38)
  (111,13.31)
  (112,13.12)
  (113,13.77)
  (114,14.02)
  (115,13.80)
  (116,13.41)
  (117,13.64)
  (118,13.49)
  (119,14.44)
  (120,14.43)
  (121,14.72)
  (122,15.29)
  (123,14.73)
  (124,14.28)
  (125,14.75)
  (126,15.62)
  (127,14.75)
  (128,14.83)
  (129,15.18)
  (130,16.04)
  (131,15.67)
  (132,15.75)
  (133,15.31)
  (134,15.60)
  (135,15.59)
  (136,15.42)
  (137,15.79)
  (138,16.00)
  (139,15.94)
  (140,16.25)
  (141,16.01)
  (142,16.50)
  (143,16.15)
  (144,15.90)
  (145,16.24)
  (146,16.91)
  (147,16.70)
  (148,16.94)
  (149,16.98)
  (150,16.85)
  (151,17.26)
  (152,17.50)
  (153,17.69)
  (154,17.41)
  (155,17.90)
  (156,17.51)
  (157,17.80)
  (158,16.95)
  (159,16.59)
  (160,17.17)
  (161,17.48)
  (162,17.71)
  (163,18.18)
  (164,17.80)
  (165,17.22)
    };

\end{axis}
\end{tikzpicture}
}
\caption{The training curve for naturally training a Neural Network compared to robust training using TRADES algorithm.}
\label{fig:training}
\end{center}
\end{figure}
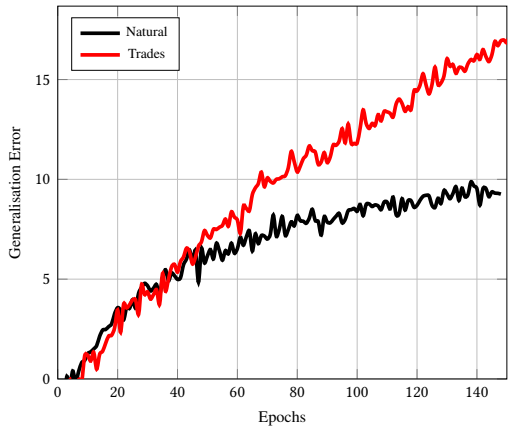

On adding adversarial noise, the goal is to estimate the extent of overfitting for different training algorithms which will indicate the impact on fault tolerance.
As shown in Figure~\ref{fig:training}, the model on training using adversarially robust algorithm overfits significantly compared to model trained using natural Stochastic Gradient Descent.
In this particular architecture, the generalization error of robust models is about 17\% compared to 9\% error of naturally trained model.

For CIFAR10 dataset, the performance for robust training using TRADES is evaluated and compared with the baseline of Tikhonov regularization.
Tikhonov regularized models are used as a baseline since, training with input noise is shown as theoretically equivalent to Tikhonov function in the objective function.
As shown in Table 4, adding adversarial noise results in increase in the overall generalization error (extent of model overfitting).

\begin{table}[!htb]
\begin{center}
\caption{Generalization Error of Adversarially Robust Models on CIFAR10.}
\renewcommand\arraystretch{1.5}
\fontsize{6.7pt}{6.7pt}\selectfont
\resizebox{\columnwidth}{!}{
\begin{tabular}{ |c|c|c|c| }
\hline
 \textbf{Algorithm} & \textbf{Training Accuracy} & \textbf{Testing Accuracy} & \textbf{Generalization Error}\\
 \hline
 \textbf{Natural} & 95.50\% & 86.24\% & 9.26\%\\
 \textbf{Tikhonov} & 89.19\% & 82.24\% & 6.95\%\\
\textbf{TRADES} & 93.26\% & 76.04\% & 17.22\% \\
 \hline
\end{tabular}
}
\end{center}
\label{generr_tab}
\end{table}


\pgfmathdeclarefunction{gauss}{2}{%
  \pgfmathparse{1/(#2*sqrt(2*pi))*exp(-((x-#1)^2)/(2*#2^2))}%
}

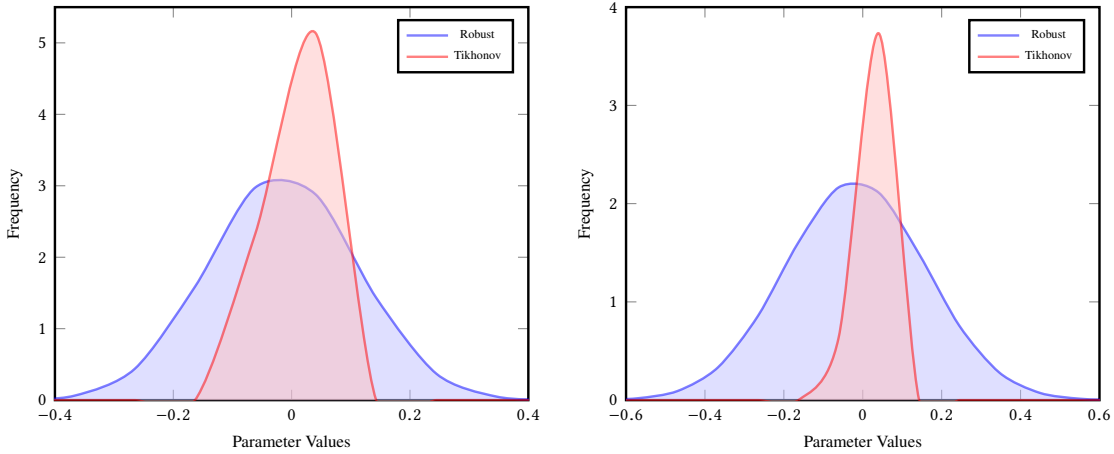
\begin{figure*}[htb!]
\begin{center}
\resizebox{1.8\columnwidth}{!}{%
\begin{tabular}{lllll}

\begin{tikzpicture}[font=\tiny]
\begin{axis}[every axis plot post/.append style={
  mark=none,domain=-2:3,samples=50,smooth},
legend style={font=\tiny},
legend pos =  north east,
line width=1.0pt,
xmin=-0.4,
xmax=0.4,
ymin=0,
ymax=5.5,
legend entries={Robust, Tikhonov},
ylabel={Frequency},
xlabel={Parameter Values}
]

\addplot[color=blue,fill=blue!25,solid,smooth,opacity=0.5] {gauss(-0.015473195638284549,0.12537815704985425)}; 
\addplot[color=red,fill=red!25,solid,smooth,opacity=0.5] {gauss(-0.002083169137470679,0.03245157647847816)};

\end{axis}
\end{tikzpicture} &

\begin{tikzpicture}
\begin{axis}[every axis plot post/.append style={
  mark=none,domain=-2:3,samples=50,smooth},
legend style={font=\tiny},
legend pos =  north east,
line width=1.0pt,
xmin=-0.6,
xmax=0.6,
ymin=0,
ymax=4,
legend entries={Robust, Tikhonov},
ylabel={Frequency},
xlabel={Parameter Values}
]

  \addplot[color=blue,fill=blue!25,solid,smooth,opacity=0.5] {gauss(-0.018198278915159495,0.17901387309432049)}; 
  \addplot[color=red,fill=red!25,solid,smooth,opacity=0.5] {gauss(-0.0004167126215661353,0.02377219502758249)}; 

\end{axis}
\end{tikzpicture}

\end{tabular}
}
\caption{\textbf{(Left) CIFAR10.}  \textbf{(Right) FashionMNIST.} For both the datasets, the distribution for the model trained using Tikhonov regularisation has a lower standard deviation of the parameter distribution compared to the models trained using adversarial input noise. This difference in the distributions indicate that training with noise enhances fault tolerance only for small values of noise and it is not equivalent to Tikhonov regularisation for adversarial noise.}
\label{fig:cifar10}
\end{center}
\end{figure*}

For the FashionMNIST dataset, the generalization error for the TRADES adversarial training algorithm is evaluated.
As shown in Table 5, the generalization error for the models trained using adversarial noise is significantly higher compared to the generalization of the Tikhonov regularization and naturally trained model without any additional optimizations.
This indicates that adversarially computed noise, in fact, has a negative impact on the fault tolerance.
On evaluating the generalization error on Multilayer Perceptron, the error increases from 3.29\% for regularized model to 8.83\% for robust models.
A similar pattern is observed for a more complex Convolutional Neural Network, with the generalization error increases from 4.90\% for a regularized model to 8.01\% for robust models.

\begin{table}[!htb]\label{fmnist_adv}
\begin{center}
\caption{Generalization Error of Adversarially Robust Models on FashionMNIST on both CNN and MLP architectures.}
\renewcommand\arraystretch{1.5}
\fontsize{6.7pt}{6.7pt}\selectfont
\resizebox{\columnwidth}{!}{
\begin{tabular}{ |c|c|c|c| }
\hline
\multicolumn{4}{|c|}{\textbf{Multilayer Perceptron Architecture}} \\
\hline
 \textbf{Algorithm} & \textbf{Training Accuracy} & \textbf{Testing Accuracy} & \textbf{Generalization Error}\\
 \hline
\textbf{Natural} & 98.83\% & 90.23\% & 8.60\%\\
\textbf{Tikhonov} & 90.99\% & 87.70\% & 3.29\%\\
\textbf{TRADES} & 98.76\% & 89.93\% & 8.83\%\\
\hline
\multicolumn{4}{|c|}{\textbf{Convolutional Neural Network Architecture}} \\
\hline
\textbf{Algorithm} & \textbf{Training Accuracy} & \textbf{Testing Accuracy} & \textbf{Generalization Error}\\
\hline
\textbf{Natural} & 98.60\% & 90.59\% & 8.01\%\\
\textbf{Tikhonov} & 95.94\% & 91.04\% & 4.90\%\\
\textbf{TRADES} & 99.59\% & 90.73\% & 8.86\%\\
\hline
\end{tabular}
}
\end{center}
\end{table}

\subsection{Comparing Fault Tolerance through Parameter Distribution}

Alternatively, another approach to estimate the fault tolerance is by evaluating the standard deviation of the trained model's parameter distribution \cite{duddu2019adversarial}.
The standard deviation of the parameters $\theta$ of the model is written as,
\begin{equation}
\sigma = \sqrt{\frac{\sum_{i} |\theta_{i} - \bar\theta|}{N}}
\end{equation}
where $\bar\theta$ is the average of all the parameter values and $N$ is the total number of parameters in the model.
Higher the standard deviation of the parameter distribution, more varied are the parameter values due to which some nodes are given more importance over the others.
Here, the loss of those important nodes in case of random faults results in a significant drop in accuracy.
For low $\sigma$, the parameters give equal weightage to all the nodes and hence, the loss of a few nodes does not impact the overall model performance.

In Figure~\ref{fig:cifar10} (left) for the CIFAR10 dataset, the standard deviation of the adversarially robust model is 0.125378 compared to 0.032451 of the regularized model parameter distribution.
For FashionMNIST dataset (Figure~\ref{fig:cifar10} (right)), the standard deviation of the parameter distribution for robust model is 0.17901 while the deviation for regularized model is 0.023772.
This indicates that the fault tolerance of models trained using adversarial noise is significantly less than Tikhonov regularized model or naturally trained model.

\subsection{Impact of Varying $\epsilon_{adv}$}

An important study is to evaluate the impact of increasing the overall range of perturbation added to the inputs, i.e, $\epsilon_{adv}$.
Increasing $\epsilon_{adv}$, results in increasing the overall noise region from which the noise can be sampled.
Thus, this results in increasing the strength of the noise added to the input.

\begin{table}[!htb]
\begin{center}
\caption{The impact of varying $\epsilon_{adv}$ values on the fault tolerance for CIFAR10 dataset.}
\renewcommand\arraystretch{1.5}
\fontsize{6.7pt}{6.7pt}\selectfont
\resizebox{\columnwidth}{!}{
\begin{tabular}{ |c|c|c|c| }
\hline
 \textbf{$\epsilon_{adv}$} & \textbf{Training Accuracy} & \textbf{Testing Accuracy} & \textbf{Generalization Error}\\
 \hline
\textbf{2/255} & 90.12\% & 81.93\% & 8.19\%\\
\textbf{4/255} & 91.02\% & 80.05\% & 10.97\%\\
\textbf{8/255} & 93.26\% & 76.04\% & 17.22\%\\
 \hline
\end{tabular}
}
\end{center}
\label{advepsilon}
\end{table}

As seen in Table 6, the overall fault tolerance measured as the difference in training and testing accuracy increases with an increase in the noise budget $\epsilon_{adv}$.
Specifically, for the case of Convolutional Neural Network trained on CIFAR10 dataset, increasing $\epsilon_{adv}$ from 2/255 to 8/255, the generalization error increases (equivalently fault tolerance decreases) from 8.19\% to 17.22\%.

\subsection{Simulation of Faults for Adversarially Robust Models}
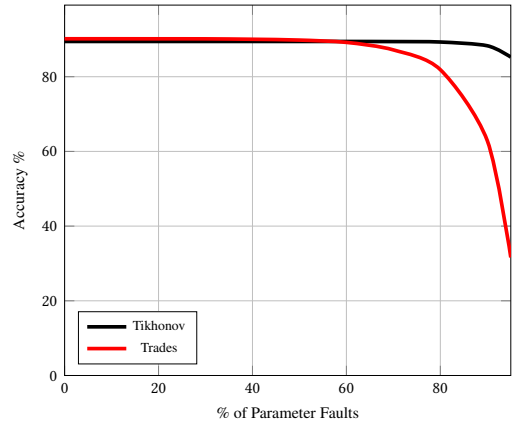
\begin{figure}[htb!]
\begin{center}
\resizebox{0.8\columnwidth}{!}{%
\begin{tikzpicture}[font=\scriptsize]
\begin{axis}[
legend style={font=\scriptsize},
legend pos =  south west,
legend entries={Tikhonov, Trades},
ylabel={Accuracy \%},
xlabel={\% of Parameter Faults},
xmin=0, ymin=0,
xmax=95,
every axis plot/.append style={ultra thick},
grid=major
]
\addplot[
    color=black,
    solid,
    smooth
    ]
    coordinates {
  (0,89.44)
  (10,89.44)
  (20,89.44)
  (30,89.44)
  (40,89.44)
  (50,89.44)
  (60,89.44)
  (70,89.41)
  (80,89.29)
  (90,88.29)
  (95,85.25)
    };
\addplot[
      color=red,
      solid,
      smooth
    ]
    coordinates {
    (0,90.16)
    (10,90.16)
    (20,90.15)
    (30,90.14)
    (40,90.03)
    (50,89.81)
    (60,89.19)
    (70,87.18)
    (80,81.88)
    (90,62.87)
    (95,31.61)
    };

\end{axis}
\end{tikzpicture}
}
\caption{Comparison of performance drop in model trained using the theoretically equivalent Tikhonov regularisation (fault tolerant) and model trained using adversarial input noise.}
\label{fig:fault}
\end{center}
\end{figure}

Since the fault model considered in the paper is random faults resulting in stuck at "0" values, these faults are simulated for adversarially robust models and compared with Tikhonov regularized models.
As seen in Figure~\ref{fig:fault}, on increasing the faults into the parameters from 50\% to 90\%, the accuracy drop in the case of the regularized model is 1.14\% compared to 26.93\% of the robust model.
This confirms the above analysis, indicating that adversarially robust models are less fault-tolerant to regularized models.

\section{Differential Privacy and Fault Tolerance}

In this section, the impact of adding noise to the gradients, to mitigate inference attacks via Differential Privacy, on fault tolerance is considered.

For the Convolutional Neural Network (Table 7), a clear trade-off between the generalization error (fault tolerance) and accuracy can be seen.
Higher fault tolerance can be achieved at the cost of low test accuracy.
This trade-off is also observed by other standard functions such as L1 and L2 regularizers, however, they do not provide privacy guarantees.
As the privacy leakage bound $\epsilon_{dp}$ is increased from 0.49 to $10^6$, the generalization error increases from 0.75\% to 4.40\%.
An increase of $\epsilon_{dp}$ indicates more information leakage.
This indicates that fault tolerance and privacy are highly correlated with each other, i.e, increasing the privacy (lowering $\epsilon_{dp}$) will also increase the overall fault tolerance at the cost of test accuracy.

\begin{table}[!htb]
\begin{center}
\caption{CNN: Generalization Error of Differentially Private Models on FashionMNIST.}
\renewcommand\arraystretch{1.5}
\fontsize{6.7pt}{6.7pt}\selectfont
\resizebox{\columnwidth}{!}{
\begin{tabular}{ |c|c|c|c| }
\hline
\textbf{} & \textbf{Training Accuracy} & \textbf{Testing Accuracy} & \textbf{Generalization Error}\\
\hline
\textbf{Natural} & 97.06\% & 89.92\% & 7.14\%\\
\textbf{Tikhonov} & 90.12\% & 89.43\% & 0.69\%\\
\hline
 \textbf{$\epsilon_{dp}$} & \textbf{Training Accuracy} & \textbf{Testing Accuracy} & \textbf{Generalization Error}\\
 \hline
\textbf{0.49} & 76.84\% & 76.09\% & 0.75\%\\
\textbf{2.97} & 84.35\% & 83.22\% & 1.13\%\\
\textbf{24.66} & 87.72\% & 86.26\% & 1.46\%\\
\textbf{2$\times\textbf{10}^6$} & 94.21\% & 89.81\% & 4.40\% \\
 \hline
\end{tabular}
}
\end{center}
\label{dp_cnn}
\end{table}

As seen in Table 8 for MLP based model, a similar pattern is observed where the generalization error increases from 1.01\% to 8.29\% as the values of $\epsilon_{dp}$ increases.

\begin{table}[!htb]
\begin{center}
\caption{MLP: Generalization Error of Differentially Private Models on FashionMNIST.}
\renewcommand\arraystretch{1.5}
\fontsize{6.7pt}{6.7pt}\selectfont
\resizebox{\columnwidth}{!}{
\begin{tabular}{ |c|c|c|c| }
\hline
\textbf{} & \textbf{Training Accuracy} & \textbf{Testing Accuracy} & \textbf{Generalization Error}\\
\hline
\textbf{Natural} & 99.58\% & 89.60\% & 9.98\%\\
\textbf{Tikhonov} & 88.96\% & 87.45\% & 1.51\%\\
\hline
\textbf{$\epsilon_{dp}$} & \textbf{Training Accuracy} & \textbf{Testing Accuracy} & \textbf{Generalization Error}\\
\hline
\textbf{0.49} & 80.44\% & 79.43\% & 1.01\%\\
\textbf{2.97} & 85.70\% & 83.64\% & 2.06\%\\
\textbf{24.66} & 88.21\% & 85.43\% & 2.78\%\\
\textbf{2$\times\textbf{10}^6$} & 96.64\% & 88.35\% & 8.29\%\\
 \hline
\end{tabular}
}
\end{center}
\label{dp_mlp}
\end{table}

\begin{mythm}
Given a Machine Learning Model trained using ($\epsilon_{dp},\delta_{dp}$)-Differential Privacy, the model's fault tolerance metric, given by the generalization error, is bounded by $e^{\epsilon_{dp}} -1 + \delta_{dp}$.
\end{mythm}

\textbf{Proof Sketch.} Differential Privacy is a strong notion of stability where the change in the data point in the training data should not change the final output.
Further, fault tolerance is also a notion of stability where a change in the model architecture should not change the final output.
A Differentially private mechanism is also uniform RO stable and the generalization error of the mechanism can be bounded by $e^{\epsilon_{dp}}-1+\delta_{dp}$ \cite{JMLR:v17:15-313}.
Since generalization error is used to measure the relative fault tolerance between different models, the corresponding fault tolerance is bounded by $e^{\epsilon_{dp}}-1+\delta_{dp}$.

\textbf{Proof.} Given the data population $P$ of all possible input and output pairs, the model is trained on a subset of data $D_{train}$ sampled from P by minimising the training error,
\begin{equation}
E_{train} = \frac{1}{n_{train}} \sum_{i} l(f(\theta,x_i),y_i)
\end{equation}

In order to evaluate the performance on any possible sample that the model might encounter, we evaluate the error on the testing dataset $D_{test}$ sampled from $P$, where $D_{test} \cap D_{train} = \phi$.
\begin{equation}
E_{test} = \frac{1}{n_{test}} \sum_{i'}l(f(\theta,x_{i'}),y_{i'})
\end{equation}
The generalization error is given by the difference between the testing ($E_{test}$) and training error ($E_{train}$).

A mechanism which satisfies ($\epsilon_{dp},\delta_{dp}$)-Differential Privacy also satisfies uniform RO stability \cite{JMLR:v17:15-313}.
Hence, for datasets D and D' differing by a single data point,

\begin{equation}
|E_{D} - E_{D'}| \leq e^{\epsilon_{dp}}-1+\delta_{dp}
\end{equation}

Further, generalizing this result for the training dataset and testing dataset,

\begin{equation}
|E_{train} - E_{test}| \leq e^{\epsilon_{dp}}-1+\delta_{dp}
\end{equation}

Since the fault tolerance is measured as the difference in the training and testing error, we can see that this is bounded by $e^{\epsilon_{dp}}-1+\delta_{dp}$ on training the model with ($\epsilon_{dp},\delta_{dp}$)-Differential Privacy.
This result on provable bound on generalization error is based on the folklore theorem by Frank McSherry.
For small values of $\epsilon_{dp}$, $e^{\epsilon_{dp}} \approx 1 + \epsilon_{dp}$ and hence, $e^{\epsilon_{dp}}-1+\delta_{dp}$ can be written as $\epsilon_{dp} + \delta_{dp}$ which is agreement with folklore theorem.
Hence, training for privacy objective using Differential privacy provides an alternate approach for enhancing fault tolerance with a provable bound on the generalization error.

\section{Conclusions}\label{conclusions}

Designing a trustworthy Machine Learning system requires to understand the trade-offs between different aspects of trust.
This work highlights the trade-offs between three such aspects of trust in Machine Learning, namely, reliability, privacy, and adversarial robustness.
This work considers two adversarial settings, with a security threat model where the adversary aims to force the model to misclassify by adding adversarial noise to the input, and a privacy threat model where the adversary aims to infer whether a data point was part of the sensitive training data or not.
Under the security threat model, the impact of fault tolerance on adversarially robust Neural Networks is evaluated and robust Neural Networks are observed to have lower the fault tolerance due to overfitting.
Under the privacy threat model, it is shown that Differentially Private models exhibit fault tolerance for a careful choice of privacy parameters ($\epsilon_{dp},\delta_{dp}$).
Hence, fault tolerance can be achieved by training models with privacy objective.
Theoretically, the bound on the model's generalization error is shown in terms of the parameters for Differential Privacy.
This study is a crucial step towards understanding the design of trustworthy Machine Learning systems.

\section*{Acknowledgement}

Valentina E. Balas would like to thank the European Research Development Fund under the Competitiveness Operational Program (BioCell-NanoART = Novel Bio-inspired Cellular Nano-architectures, POC-A1-A1.1.4-E nr. 30/2016) for supporting the research.

\bibliographystyle{ACM-Reference-Format}
\bibliography{paper}

\end{document}